# The Physics Case for Extended Tevatron Running

D. R. Wood
*Northeastern University, Boston, MA 02115 USA*
On behalf of the CDF and D0 Collaborations

Run II of the Tevatron collider at Fermilab is currently scheduled to end late in 2011. Given the current performance of the collider and of the CDF and DØ detectors, it is estimated that the current data set could be approximately doubled with a run extended into 2014. A few examples are presented of the physics potential of these additional statistics. These are discussed in the context of the expected reach of the LHC 7 TeV data and the existing Tevatron data. In particular, an extraordinary opportunity is described which could probe the existence of a standard model Higgs boson with mass in the currently preferred region between 115 GeV and 150 GeV.

## 1. CONTEXT

The Tevatron accelerator has been running extremely well and has delivered about 2 fb$^{-1}$ per year of integrated luminosity to the CDF and DØ experiments over the last three years. No significant upgrades are planned in the future, but based on projections of recent performance, the Tevatron would deliver about 2.5 fb$^{-1}$ per year in the coming years. The Tevatron run is currently approved to continue only through September 2011. If the run is extended by three years to 2014, a total integrated luminosity of about 19 fb$^{-1}$ is anticipated. After data taking efficiency and quality requirements are taken into account, this should yield about 16 fb$^{-1}$ appearing in final analyses, which corresponds to approximately double the data available today. Many analyses at the Tevatron are still limited by statistics, so these would obviously benefit from a doubling of the integrated luminosity. To evaluate these benefits in context, however, one must also take into account the expected performance of the CDF and DØ detectors, and the concurrent operation of the LHC at CERN which will be exploring some of the same physics with higher energy collisions and lower integrated luminosity.

## 2. COMPLEMENTARITY

In many physics topics, the data from the Tevatron and the LHC play complementary roles due to the differences between the two machines in collision energy (2 TeV vs. 7 TeV), integrated luminosity (~20 fb$^{-1}$ vs. 1-2 fb$^{-1}$) and initial state particles (proton-antiproton vs. proton-proton). For example, the two colliders tend to probe different regions of the parton distributions within the proton with the LHC processes predominantly originating from lower values of Feynman-x. For Feynman-x values above 0.05, the two-jet production cross section at the Tevatron is more than 100 times that at the LHC. The types of initial state partons also vary. For example, at the Tevatron, $t\bar{t}$ production is dominated by the $q\bar{q}$ initial state, while at the LHC it is dominated by gluon-gluon fusion. In addition, the proton-antiproton initial state is CP-symmetric, which facilitates sensitive tests of CP violation and provides a meaningful convention for measuring forward-backward asymmetries. An interesting example is illustrated in Fig. 1, which shows recent distributions from CDF[1] and DØ[2] of the forward-backward charge asymmetry in $t\bar{t}$ production, both of which show some excess in the forward direction. This is best pursued further with additional Tevatron data; such an asymmetry has no straightforward equivalent in a proton-proton machine like the LHC.

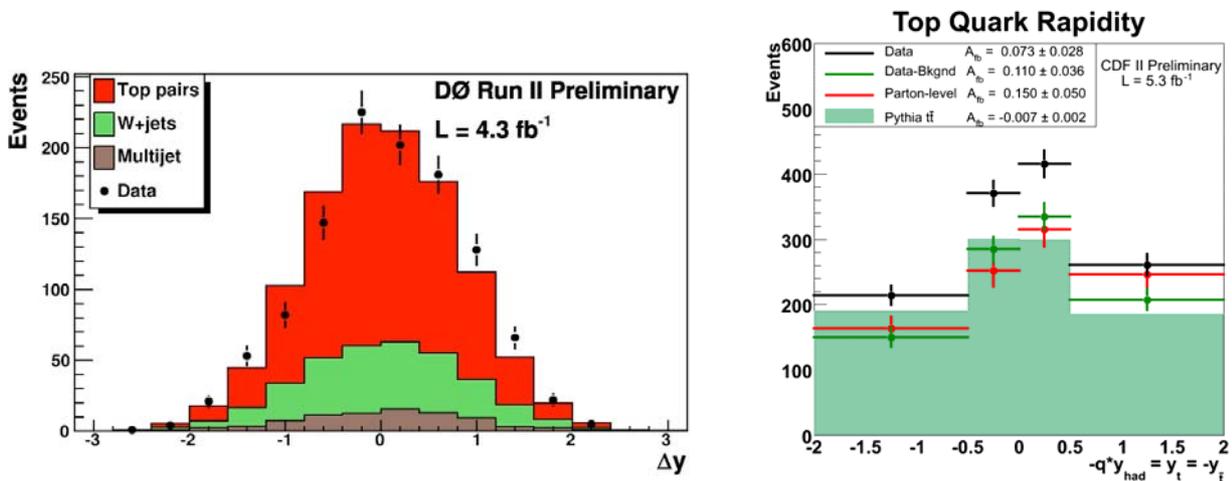

Figure 1: The charge-signed rapidity difference distribution for $t\bar{t}$ production from DØ and CDF. Both sets of data show a shift to positive values with respect to standard model predictions.

## 3. LEGACY MEASUREMENTS

Some precision measurements gain no particular advantage from the higher energy of the LHC, and at the Tevatron they benefit from years of careful study of systematic effects. Thus the Tevatron results may stand for years as the most precise results. The most important measurements in this category are those of the top quark mass and the W boson mass. The precise knowledge of these masses, combined within the context of the standard model (SM), predicts the value of $m_H$, the mass of the Higgs boson. With a future direct measurement of $m_H$, the uncertainties on the masses of the top quark and W boson would limit the power of the critical self-consistency test of the model of electroweak symmetry breaking.

The precision on the world-average W boson mass is 23 MeV. The strongest contribution to this comes from the Tevatron combination [3], with its uncertainty of 31 MeV, or 0.04%. Since many of the systematic uncertainties are controlled with the Z→ℓℓ sample that is accumulated along with the W→ℓν sample, the systematic precision improves with additional data at the Tevatron. A goal for the full data sample is a precision of 15 MeV on the world average.

Figure 2 shows measured and projected top quark mass uncertainties from the Tevatron. The current uncertainty is 1.1 GeV[4], and it is still limited by statistics. The upper range of the shaded projection region assumes no improvement in systematic uncertainties apart from those from the jet energy scale, which is measured in-situ in the top quark data sample. The lower band assumes additional progress in reducing some of the other systematic uncertainties.

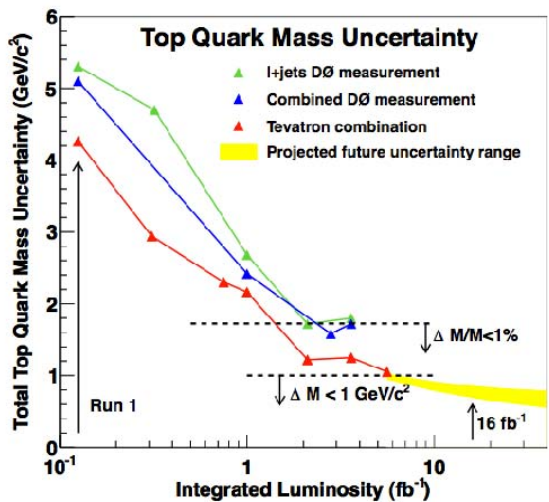

Figure 2: The uncertainty on the top quark mass from the DØ experiment and from combined DØ and CDF results. The triangles show actual measurements, and the shaded band shows projections to future measurements.

If we assume these future uncertainties of 1 GeV on the top quark mass and 15 MeV on the W boson mass, and assume no change in central values of these masses, the SM constraints would give a bound of $m_H$<117 GeV at the 95% CL. With the existing LEP2 exclusion below 114 GeV, this would leave a very narrow allowed region for $m_H$. If the direct searches exclude this region or discover the Higgs boson outside of this range, this would then provide an essential indication of a non-standard model nature of the Higgs boson.

## 4. HINTS AND EXCESSES

In the existing Tevatron data set, there are several places where disagreements with the SM are seen. None of these is significant enough to establish a discovery, but they could be significant if they remain prominent with a doubling of the data set. The disagreements include the $t\bar{t}$ forward-backward charge asymmetry mentioned earlier. The direct searches for a fourth generation quark, t-prime, also show an excess corresponding to a t-prime mass around 400 GeV, both at CDF[5] and DØ[6].

The $B_s$ system has so far been studied almost exclusively at the Tevatron. The greatest focus of this effort has been on understanding CP-violation in the $B_s$ system. Two particular results have shown some tension with the SM. The measurement of the CP-violating phase angle, $\varphi_s$, is particularly interesting. The most recent Tevatron combination of the $\varphi_s$ measurement in $B_s \rightarrow J/\psi\varphi$ [7] shows a deviation from the SM of about 2-σ. In addition, DØ has measured the charge asymmetry in like-sign dimuons [8], which is sensitive to a linear combination of $a_{sl}^s$ and $a_{sl}^d$, the CP-asymmetries in the semileptonic decays of $B_s$ and $B_d$, respectively, as shown in Fig. 3. The LHCb experiment will be able to examine the CP asymmetries of the $B_s$ system, but their planned semileptonic decay asymmetry analysis [9] probes a different linear combination of $a_{sl}^s$ and $a_{sl}^d$. As shown in Fig. 3, the expected limits from LHCb are nearly orthogonal to the constraints from the dimuon charge asymmetry at DØ, so further Tevatron data will again be highly complementary to the LHC.

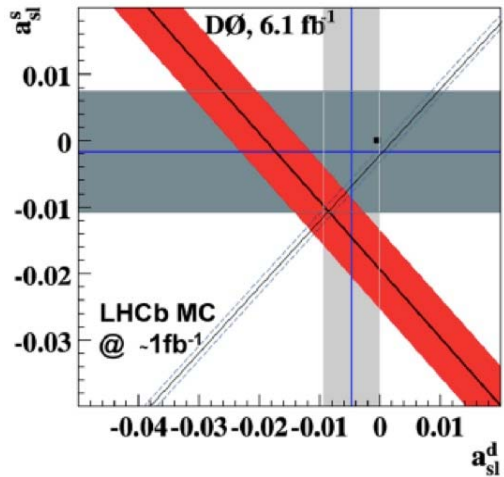

Figure 3: Existing and projected constraints on the $a^s_{sl}$ and $a^d_{sl}$, the semileptonic decay asymmetries of $B_s$ and $B_d$. The horizontal band shows the constraint from the DØ $B_s$ semileptonic analysis, and the vertical band shows the world average $B_d$ semileptonic constraints, coming from the B-factories. The wider diagonal band is the current constraint from the DØ dimuon charge asymmetry, while the narrow orthogonal band is the projected sensitivity of the inclusive semileptonic asymmetry analysis of LHCb.

## 5. STANDARD MODEL HIGGS

The understanding of electroweak symmetry breaking is one of the central questions of particle physics today. The simplest viable realization of this symmetry breaking requires the existence of a yet-undiscovered Higgs scalar. In recent years, the Tevatron experiments have made remarkable progress in the search for the Higgs boson, and the continuation of this search is the dominant motivation for extending the Tevatron run. The Tevatron has already excluded a high-mass region for the Higgs boson between 158 and 175 GeV, where its decay is dominantly H→WW. It has also made substantial progress in the lower mass region, where $H \to b\bar{b}$ dominates, via the associated production modes, $p\bar{p} \to WH + X$ and $p\bar{p} \to ZH + X$. The log-likelihood ratio plot in Fig. 4 gives a compact summary of the current state of the Higgs boson search at the Tevatron [10]. At high masses, around $m_H$=165 GeV, the data are in very good agreement with the background-only predictions while they deviate significantly from the signal prediction. At lower masses (115<$m_H$<150 GeV), the data are still compatible with both the signal and the background hypotheses, but the two hypotheses are now separated by more than 1-σ and this separation will increase with more integrated luminosity.

The prospects for Higgs sensitivity with future data at the Tevatron are very exciting. Figure 5 shows the expected luminosity required for 95% CL exclusion and for 3-σ evidence as a function of $m_H$. Along with additional statistics, extended running of the Tevatron would also be accompanied by detector aging effects as well as improvements to the analysis, and these are all taken into account in the projections. The improvements include increases in lepton identification efficiency and acceptance, inclusion of additional production and decay modes, refinement of the jet energy resolution, and more effective multivariate discrimination techniques. The difference between the central lines and the upper edges of the band shows the anticipated additional luminosity required to overcome the losses due to detector aging. The difference between the upper and lower edge of the bands shows the gain in equivalent luminosity expected from incorporating the analysis improvements. The lower edge of the band represents the best estimate of expected sensitivity including both aging and improvements. Note that with 16 fb$^{-1}$ of data: (1) the Higgs boson will be strongly excluded over the entire mass range below 200 GeV if it does not indeed exist; (2) 3-σ or greater evidence is expected if the Higgs exists in the preferred mass region, and (3) approximately 4-σ evidence is expected if $m_H$ is around 115 GeV.

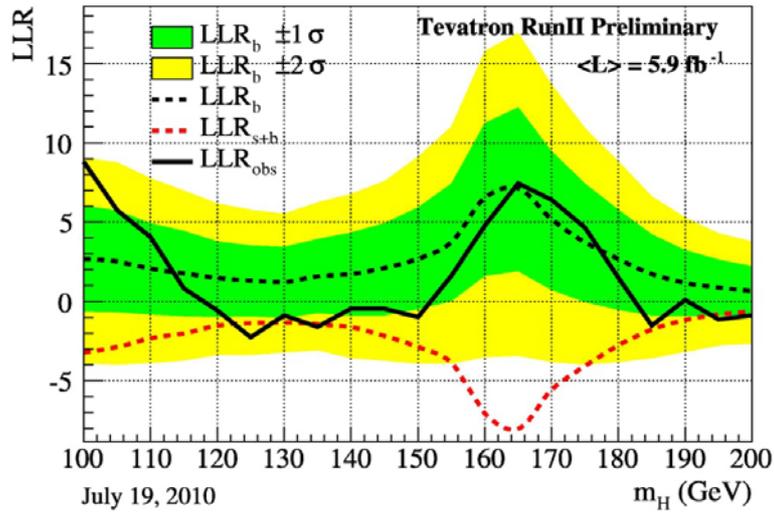

Figure 4: The logarithm of the likelihood ratio (LLR) of background-only to signal-plus-background hypotheses for the combined Tevatron data, as a function of Higgs boson mass. The upper dashed curve shows the prediction for the background-only hypothesis, and the shaded bands show the expected 1-σ and 2-σ variations. The lower dashed curve shows the expectation of the signal hypothesis and the solid curve shows the observed LLR.

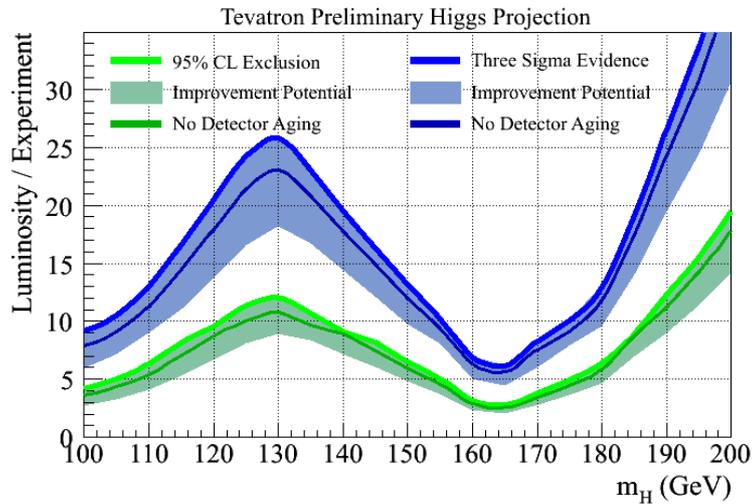

Figure 5: Projections of the luminosity required to establish 95% CL exclusion (lower band) or 3-σ evidence (upper band) at the Tevatron, as a function of Higgs boson mass. See text for details.

If the SM Higgs boson exits, the LHC should eventually be able to establish rigorous discovery for any reasonable value of $m_H$. With 1fb$^{-1}$ of 7 TeV data, the LHC should quickly be able to confirm or refute the high mass exclusion from the Tevatron via the H→WW mode, but sensitivity to a Higgs Boson in the lower end of the mass range requires more time and luminosity. In addition, the larger backgrounds from $b\bar{b}$, W+jets, and $t\bar{t}$ at the LHC make it difficult to examine the $H \to b\bar{b}$ mode; the expected discovery mode for a low-mass Higgs at the LHC is the rare decay H→γγ with some contribution from H→ττ.

For a low-mass Higgs boson, information that would come from the Tevatron and the LHC would be highly complementary and enormously valuable in understanding the Higgs sector, especially if it turns out not to be the simplest SM variety. The Tevatron data would not only give an early indication of $m_H$, they would give information about couplings and decays that would be very difficult to extract at the LHC. The Tevatron signals would probe both the coupling to b-quarks through the Higgs decay and the coupling to W and Z bosons, since the associated production mode is used. While the loop-induced H→γγ mode at the LHC could provide an excellent measurement of $m_H$, this mode is sensitive to beyond-the-SM effects that could reduce its rate. These effects would have little influence on the $H \to b\bar{b}$ decay rate.

In summary, an extension to the Tevatron run could provide an opportunity for early evidence of the Higgs boson and would probe production and decay modes that are highly complementary to those studied at the LHC. In particular, if the simple SM scenario is not realized in nature, the comparison of LHC and Tevatron data could be critical in understanding the mechanism of electroweak symmetry breaking.

## Ackowledgments

The support of the D0 and CDF collaborations is gratefully acknowledged. The author's work is supported by NSF grant PHY-0757561.